\documentclass[12pt]{article}
\usepackage{graphicx}
\catcode`\@=11
\catcode`\@=12
\def\com#1#2{\Big[#1,#2\Big ]}
\newcommand{\bea}{\begin{eqnarray}}
\newcommand{\eea}{\end{eqnarray}}
\newcommand{\hp}{\hat{\Phi}}
\newcommand{\hA}{\hat{A}}

\def\be{\begin{equation}}
\def\ee{\end{equation}}

\def\fr{\frac}
\def\a{\alpha}
\def\b{\beta}

\def\d{\delta}
\def\e{\epsilon}

\def\l{\lambda}
\def\m{\mu}
\def\n{\nu}
\def\n{\nu}
\def\r{\rho}
\def\s{\sigma}

\def\W{\Omega}

\def\d{\delta}

\def\L{\Lambda}

\def\p{\partial}
\def\t{\tilde}

\def\nn{\noindent}
\def\no{\nonumber}

\def\cA{{\cal A}}  
\def\cD{{\cal D}}

  \def\cO{{\cal O}}
  
\def\cS{{\cal S}}  
 \def\cW{{\cal W}}

\begin{document}
\title{{\small\hfill IMSc/2004/06/26}\\
Non-commutative Duality: High Spin Fields and $CP^1$ Model with Hopf Term}
\author{T. R. Govindarajan,\thanks{trg@imsc.res.in}\\
The Institute of Mathematical Sciences,\\
C I T Campus, Taramani, Chennai, 600 113, India.\\
\em{and}\\
E. Harikumar,\thanks{hari@fma.if.usp.br}\\
Instituto de F\'{\i}sica, Universidade de S\~{a}o Paulo,\\
Caixa Postal 66318, 05315-970, S\~{a}o Paulo - SP, Brazil}
\maketitle
\begin{abstract}  

We show that the non-commutative $CP^1$ model coupled with Hopf term
in 3 dimensions is equivalent to an interacting spin-$s$ theory where
the spin $s$ of the dual theory is related to the coefficient of the
Hopf term. We use the Seiberg-Witten map in studying this
non-commutative duality equivalence, keeping terms to order $\theta$
and show that the spin of the dual theory do not get any
$\theta$ dependant corrections. The map between current correlators
show that topological index of the solitons in the non-commutative
$CP^1$ model is unaffected by $\theta$ where as the Noether charge of
the corresponding dual particle do get a $\theta$ dependence. We also
show that this dual theory smoothly goes to the limit $\theta\to 0$
giving dual theory in the commutative plane.

\end{abstract}
\medskip
\nn {PACS Numbers:11.10.Nx,11.10.Ef}\\ 
\nn Keywords: non-commutative plane, duality, high spin fields \\
\newpage
\section{Introduction}\label{intro}

The recent developments in non-commutative(NC) geometry\cite{con} and
string theory\cite{sw} has motivated the study of different features
of field theory models constructed on NC space-time\cite{rev}.  The
non-commutativity of the space-time introduces non-linear and
non-local effect and hence the field theory models constructed on such
spaces have many interesting features which their commutative
counterparts do not share, like the possibility of novel soliton
solutions \cite{nov}, UV/IR mixing \cite{min} etc. The UV/IR mixing
which is a characteristic feature of non-commutative(NC) field
theories affect their renormalisability \cite{min,iya}. Recently
renormalisabilty of super-symmetric field theory models in NC setting
have also been studied and it has been argued that the non-commutative
(NC) super-symmetric gauge theories have better
renormalisabilty\cite{riv}. NC super-symmetric quantum mechanical
models \cite{us} have also been constructed and studied. Recently
fermionic field theory models has been studied on NC space-time which
avoids the Fermion doubling problem and serves as alternative to
Lattice regularisation\cite{bal1}. Quantum theories with space-time
non-commutativity have also been considered recently with potential
applications \cite{bal2}.

Seiberg-Witten (SW) map \cite{sw} allows to re-express the NC gauge
theoretic models in terms of the ordinary gauge fields and the NC
parameter $\theta$ and has been employed to study various aspects of NC
field theoretic models\cite{sw1,gs}. The SW map is derived by
demanding that the ordinary gauge fields which are connected by a
gauge transformation are mapped to NC fields which are likewise
related by the corresponding NC gauge transformation and this map
smoothly reduces to the commutative limit when $\theta\to 0$. Using SW
map, it has been shown recently that the NC Chern-Simons term get
mapped to standard Chern-Simons term in the commutative
plane\cite{gs}. It has been argued that the commutative limit(
i.e.,$\theta\to 0$) of NC models may not be
smooth\cite{min,lim}. Therefore it is of interest to see how some of
the well established field theoretic notions in the commutative spaces
generalises to NC settings. In this paper we investigate one such
problem, namely the dualisation of $CP^1$ model with Hopf term in NC
plane.

Study of the duality between bosonic and fermionic theories in
commutative spaces has a long history. In \cite{col} the equivalence
between Sine-Gorden and Massive Thirring model in $1+1$ dimension has
been studied. Following \cite{poly}, boson-fermion transmutation $2+1$
dimensional field theoretic models were studied in \cite{bose, trg}
and also perturbatively in \cite{des}. In \cite{trg} it has been shown
that the non-linear sigma model when coupled to Hopf term (written in
$CP^1$ language) is equivalent to an interacting spin-$s$
($s=\fr{1}{2},1,..)$ theory and the mapping between the dual fields
has been obtained. Duality and bosonisation of non-linear and
non-abelian theories has also been studied recently\cite{na,nab}.

The duality between Maxwell-Chern-Simons theory and Self-dual model in
$2+1$ dim.\cite{sd} (which is a crucial ingredient in obtaining the
'bosonisation' rules for massive Thirring model in $2+1$ dim.) has been
recently analysed in the NC settings \cite{dayi} using SW map to the
order $\theta$. Following this it has been shown that the equivalence
between the massive Thirring model and Maxwell-Chern-Simons theory(to
the {\it leading} order in the inverse fermion mass) is (not) valid in
the NC space where as the $1+1$ dimensional bosonisation is intact in
NC settings\cite{sg}. The study of NC duality and bosonisation is also
of interest as these studies can shed further light to the similar
problems in the non-abelian gauge theories since later have a similar
gauge structure as NC gauge theories. In this paper we study the
dualisation of NC $CP^1$ model coupled with Hopf term. The $CP^1$
model in NC plane has been studied and soliton solutions were obtained
recently. It has been argued that the equivalence of non-linear sigma
model and $CP^1$ model in the commutative plane do not hold good in
the NC settings \cite{cp1}.

In this paper we show that the NC $CP^1$ model coupled with Hopf term
is equivalent to NC spin-$s$ theory. We obtain this duality
equivalence using the path integral method developed \cite{bose,trg}
in implementing the approach of \cite{poly} in $2+1$ dimensional field
theoretic models. We apply this method, after re-expressing the NC
$CP^1$ model coupled with Hopf term in terms of the commutative fields
and NC parameter $\theta$ using SW map. We obtain the dual interacting
spin-$s$ theory where the spin-$s$ is given by $s=\fr{\pi}{2\l}$ where
$\l$ is the coefficient of the Hopf term. Here we obtain exact duality
equivalence between NC $CP^1$ model coupled with Hopf term and NC
spin-$s$ theory. We also obtain the mapping between the current
correlators of these two equivalent NC models.

\section{NC $CP^1$ model and SW map}

The $CP^1$ model in commutative plane is described by the action
\be
S=\int d^3 x |(\p_\mu\Phi_{a}-iA_\m\Phi_{a})|^2,~~~~a=1,2
\ee
where the complex doublet field $\Phi_a$ satisfy the conditions
\bea
|\Phi_1|^2+|\Phi_2|^2&=&4g^2\label{mod}\\
-i{\Phi}_{a}^*\p_\m\Phi_a&=&4g^2 A_\m.
\label{gc}
\eea
It has been shown that the above local $U(1)$ invariant action when coupled
to the Hopf term
\be
H=-\fr{i\l}{4\pi^2}\int d^3 x \e_{\m\n\l}A_\m\p_\n A_\l,
\ee
is equivalent to spin-$s$ theory\cite{trg}. Here the spin$-s$ is
related to the coupling strength $\l$ of the Hopf term. In this paper we
investigate this equivalence in the NC plane. The NC space-time is
defined by the co-ordinates obeying 
\be
{\com{X_\m}{X_\n}}_*=i\theta_{\m\n}
\ee
where the $*$ product is defined as
\be
{f}(x)*{ g}(x)=e^{\fr{i}{2}\theta^{ij}\partial_{i}^{x}\partial_{j}^{y}}
f(x)g(y)|_{x=y}\,.
\label{startp}
\ee
In the following we take the anti-symmetric tensor $\theta_{\m\n}$ to
be a constant.

We  start with the NC $CP^1$ model action coupled
to the  Hopf term
\be
{\hat\cS}=\int d^3 x \left[ ({\hat D}_\mu\hp_a)^\dagger ({\hat D}_\mu\hp_a)
-\fr{i\l}{4\pi^2}\e_{\m\n\l}({\hA}_\mu\p_\n
{\hA}_\l+\fr{2i}{3}{\hA}_\m {\hA}_
\n{\hA}_\l)\right]
\label{cp1act}
\ee
where the covariant derivative is defined as ${\hat D}_\mu\hp=
\p_\m\hp -i{\hA}_\m\hp$ and all the products in the above are $*$
products. This action is invariant under the NC $U(1)$
transformations
\bea
&\hp\rightarrow {\hat U}\hp&\no\\
&{\hA}_\m\rightarrow {\hat U}{\hA}_\m{\hat U}^\dagger-i\p_\m{\hat
U}{\hat U}^\dagger&
\eea
We re-express this action in terms of the commutative fields and the
non-commutative parameter $\theta$ using Seiberg-Witten(SW) map. The SW
map for the complex scalar field and the gauge field, to the order
$\theta$ are given by
\bea
&\hp=\Phi-\fr{1}{2}\theta_{\m\n}A_\m\p_\n\Phi,&\\
&{\hA}_\m=A_\m-\fr{1}{2}\theta_{\n\l}A_\n(\p_\l A_\m+F_{\l\m})&
\eea
respectively. 

Since the NC Chern-Simons term get mapped to the standard
Chern-Simons term in the commutative plane under the SW map, all the $\theta$
dependant terms comes from the first term when we apply SW map to the
action in Eqn. (\ref{cp1act}). To the order $\theta$, the SW mapped
action is
\bea
{S}&=&\int d^3x |D_\m\Phi_a|^2-h_{\m\n}(D_\n\Phi_a)^{*}(D_\m\Phi_a)
-\fr{i\l}{4\pi^2}\e_{\m\n\l}A_\mu\p_\n A_\l\label{swcpact}\\
{\rm ~~~~where~~~~}h_{\m\n}&=&\fr{1}{2}\left(\theta_{\m\a}F_{\a\n}+\theta_{\m\a}F_{\a\m}+\fr{1}{2}\eta_{\m\n}\theta_{\a\b}F_{\a\b}\right)
\label{swcp}
\eea
The second term in the action above is the new $\theta$ dependant
interaction terms introduced by the non-commutative nature of the 
space-time. The partition function for this theory is
\be
Z=\int D\a D\eta DA D\Phi_{a}^*D\Phi_a ~e^{-S}\label{pf}
\ee
where 
\bea
{S}&=&\int d^3 x \left[
|\p_\m\Phi_a|^2 -4g^2 A_{\m}^2
+h_{\m\n}(\Phi_{a}^*\p_\m\p_\n\Phi_a+4g^2A_\m A_\m)
-\fr{i\l}{4\pi^2}\e_{\m\n\l}A_\mu\p_\n A_\l\right.\no\\
&-&\left.\a_\m(4g^2
A_\m+i\Phi_{a}^*\p_\m\Phi_{a})+\eta^2-2i\eta{\sqrt{\r}}(|\Phi_1|^2+|\Phi_2|^2-4g^2)\right].
\label{sact}
\eea
Here, notice that constraint on $CP^1$ fields in Eqn. (\ref{mod}) is
implemented in the path integral through
$\r(|\Phi_1|^2+|\Phi_2|^2-4g^2)^2$, with the parameter
\footnote{The constraints are treated as functional delta function
following our earliar work\cite{trg} and also that of Mitter 
and Ramdas\cite{mitter}} $\r\to\infty$
and this term is then linearised using an auxiliary field $\eta$. The
constraint given in Eqn. (\ref{gc}) is introduced using the multiplier
fields $\a_\m$\footnote{ All the $\theta$ dependent terms coming from
the constraint in Eqn. \ref{gc} when SW map is applied cancels when it
is plugged back into SW mapped $CP^1$ action. This justifies the use of
the commutative constraint in the SW mapped action. See \cite{sgr} for a 
detailed discussion on this aspect.}. Now introducing the fields
$b_\m$ and $a_\m$ we linearise the quadratic term in $A_\m$ and the 
Chern-Simons term (Hopf term) respectively to write the partition
function of this theory as
\be
Z=\int DC D\a D\eta DB Da DA Db D\Phi_{a}^*D\Phi_a e^{-{\cS}}\label{pfe}
\ee
where the action
\bea
{\cS}&=&
\int d^3x \Phi_{a}^*\left[-D_\m D_\m + V\right]\Phi_a-
C_{\m\n}(\Phi_{a}^*\p_\m\p_\n\Phi_a+4g^2A_\m A_\n)
+(C_{\m\n}+h_{\m\n})B_{\m\n}\no\\
&-&4g^2\a_\m A_\m +\eta^2-8ig^2\eta{\sqrt{\r}}
+4g^2\left[\fr{\a_{\m}^2}{4}+\a_\m(b_\m+ik\e_{\m\n\l}\p_\n a_\l)\right]
\no\\
&-&4g^2\left[ 2ik\e_{\m\n\l}b_\m\p_\n a_\l-k^2(\e_{\m\n\l}\p_\n a_\l)^2
+ik\e_{\m\n\l}a_\m\p_\n a_\l\right].
\label{act}
\eea 
Using the auxiliary fields $C_{\m\n}$ and $B_{\m\n}$ we have
conveniently re-expressed the above action where there is no direct coupling
between $\theta$ dependant terms and the $CP^1$ fields $\Phi_\a$.
Here the covariant derivative is defined as $D_\m=\p_\m+iW_\m$ where
the gauge field is given by
\bea
W_\m&=&b_\m +\fr{1}{2}\a_\m +ik\e_{\m\n\l}\p_\n a_\l\\
{\rm~~~~ with~~~~~~~~}k&=&-\fr{\l}{(4\pi^2)(4g^2)}{\rm ~~~and~~~} V=2i
\eta\sqrt{\r}
\label{vp}
\eea

\section{Duality Equivalence}

We now carry out the integrations over $\Phi_{a}^*$ and $\Phi_a$ in
the partition function in Eqn.(\ref{pfe}) after re-writing the action
in Eqn. (\ref{act}) as
\be
{\cS}=\int d^3 x ~\Phi_{a}^* {\cO}\Phi_a + S_{0},
\ee
where
\bea
S_{0}&=&\int d^3 x C_{\m\n}\left[i\p_\m W_\n -2\cdot 4 g^2 W_\m A_\n -4g^2 W_\m
W_\n\right] -4g^2 C_{\m\n}A_\m A_\n+
(C_{\m\n}+h_{\m\n})B_{\m\n}\no\\
&-&4g^2\a_\m A_\m +\eta^2-8ig^2\eta{\sqrt{\r}}
-4g^2\left[\fr{\a_{\m}^2}{4}+\a_\m(b_\m+ik\e_{\m\n\l}\p_\n a_\l)\right]
\no\\
&-&4g^2\left[ 2ik\e_{\m\n\l}b_\m\p_\n a_\l-k^2(\e_{\m\n\l}\p_\n a_\l)^2
+ik\e_{\m\n\l}a_\m\p_\n a_\l\right],
\label{act1}
\eea
and the operator ${\cO}$ is given by
\be
{\cO}=-(\d_{\m\n}+C_{\m\n})D_\m D_\n + V.
\label{op}
\ee
Thus the partition function reduces to
\be
Z=\int DC D\a D\eta DB Da DA Db ~~~e^{-S_{0}-2~ln~ det~{\cO}}\label{pf1}.
\ee
Using the well known propertime representation of determinant for the operator
${\cO}$ defined in Eqn. (\ref{op}), we get
\be
-2~ln~det~{\cO}=2\int_{\L^{-2}}^{\infty}~\fr{d\b}{\b}\int Dq_\m(\tau)
~e^{-\int_{0}^{\b} d\tau\left[\fr{1}{4}(\d_{\m\n}-C_{\m\n}){\dot
q}_\m{\dot q}_\n + V\right]-i\oint_C W_\m dx^\m}
\label{popdet}
\ee

Notice that the $\det {\cO}$ depends on the gauge field $W_\m$ through
the Wilson loop. Also the auxiliary field $C_{\m\n}$ appears in the
$\det$ where as there is no explicit $\theta$ dependence.

Substituting this in Eqn. (\ref{pf1}) and expanding $e^{-2ln
det{\cO}}$, we get the partition function as
\bea
&Z=\int DC D\a D\eta DB Da DA Db~
(~1+\Sigma_{i=1}^{\infty}\fr{Z_n}{n!}~)~e^{-S_0}&\label{pf2}\\&{}&
\no\\
&{\rm~~where~~~~~~~}Z_n={\Pi}_{i=1}^{\infty}2^n\int~\fr{d\b_i}{\b_i}\int Dq_{\m}^i(\tau)
~e^{-\int_{0}^{\b} d\tau\left[\fr{1}{4}(\d_{\m\n}-C_{\m\n}){\dot
q}_{\m}^i{\dot q}_{\n}^i + V\right]
-i\oint_{C_i} W_\m dx^\m}.&\label{hk}
\eea
Here we notice that all the dependence of the the partition function on
the NC parameter $\theta$ comes through $S_0$.

Notice that the term $(~1+\Sigma_{i=1}^{\infty}\fr{Z_n}{n!}~)$ in
Eqn. (\ref{pf2}) above contains {\it all} the terms in the series
expansion of $e^{-2ln det{\cO}}$. We do not neglect any terms here
and thus we are evaluating the partition function {\it exactly}.
$Z_n$ in the above can be taken as the defining the paths of $\Phi_a$
particles\cite{trg}.

We consider the the first term in Eqn. (\ref{pf2})
$$Z_0=\int DC D\a D\eta DB Da DA Db~e^{-S_0}$$
which after the integrations over $b_\m,~A_\mu$ and $a_\mu$ becomes
\be
Z_0 =\int DC D\a D\eta DB Dv_\m~e^{-S_{eff}},\label{z0}
\ee
where the effective action is 
\bea
S_{eff}&=&\int d^3 x ~4g^2
\left[\fr{\a_{\m}^2}{4}+\fr{\a_{\m}F_{\m}(\theta)}{2\cdot
4g^2} +\fr{1}{4}\left(\a_\m+\fr{F_{\m}(\theta)}{4g^2}\right) C_{\m\n}^{-1}
\left(\a_\n+\fr{F_{\n}(\theta)}{4g^2}\right)~\right]\no\\
&+&\fr{1}{4\cdot{4g^2}}F_{\m}^2(\theta)
+\fr{i}{4\cdot 4g^2 k}F_\m(\theta)d_{\m\n}^{-1}F_\n(\theta)
+C_{\m\n}B_{\m\n}+\eta^2+i8g^2\eta\sqrt{\r}\no\\
&+&\fr{i4g^2}{2}C_{\m\n}\p_\m\a_\n
+g^2(\p_\m C_{\m\n})^2 -\fr{1}{k}\p_\n C_{\m\n}
 d_{\m\a}^{-1}\left[ig^2\p_\b C_{\b\a}-\fr{1}{2}F_{\a}(\theta)\right]
+\fr{1}{3}v_{\m}^2 .
\label{seff}
\eea
In the above we have used the definition
\be
F^{\m}(\theta)=\left[\theta^{\r\a}\p^\a B^{\r\m}-\theta^{\r\m}\p^\a
B^{\r\a}+\fr{1}{2}\theta^{\a\m}\p^\a B_{\s\s}\right],
\label{ft}
\ee
and $d_{\m\n}=-\e_{\m\n\l}\p_\l$. Also we use
$C_{\m\n}^{-1}$ where $C_{\m\n}C_{\n\l}^{-1}=\d_{\m\l}$. In
Eqn. (\ref{z0}) we have introduced the a new field $v_\m$ in the
measure and a Gaussian factor in the action (see
Eqn. \ref{seff}). This is done for later convenience (see
Eqns. (\ref{dpf1}) below). Thus the $Z_0$ in Eqn. (\ref{z0}) contains
the contribution from the first term in the series expansion of
$~e^{-2ln det{\cO}}$.  Next we evaluate the contribution to the
partition function from the remaining terms of this series. From
Eqns. (\ref{pf2}) and (\ref{hk}), we see that these terms contains
expectation value of the products of Wilson loops(for every $i$ in
Eqn. (\ref{hk}) we have a Wilson loop to be averaged with weight
factor $S_0$.). Here we use the fact that the averaging over the
products of Wilson loops is factorisable and hence it is equal to the
product of the averaging over the Wilson loops when the coefficient of
the Hopf term $\l=\fr{\pi}{2s}$. That is, we use the
property of the expectation value of Wilson loop $W(C_i)$,
\be
<W(C_1)....W(C_n)>=\prod_{i=1}^n <W(C_i)>
\ee
when $\l=\fr{\pi}{2s}$, which can be easily verified in a straight
forward manner in the present case by considering that the Wilson loops
are non-intersecting\cite{trg}.  Also notice that the product of
Wilson loop is nothing but the union of the Wilson loops. Using these
results we get the second term in Eqn. (\ref{pf2}) to be
\be
Z^\prime=\int D\W~\left[
{\Pi}_{i=1}^{\infty}2^n\int~\fr{d\b_i}{\b_i}\int Dq_{\m}^i(\tau)
~e^{-\int_{0}^{\b} d\tau\left(\fr{1}{4}(\d_{\m\n}-C_{\m\n}){\dot
q}_{\m}^i{\dot q}_{\n}^i + V\right)}\right]
~e^{-i\oint_{C} W_\m dx^\m -S_0}
\ee
where the measure $D\W= DC D\a D\eta DB Da DA Db$.

Now we carry out the integrations over the fields $b$ and $A$. Here
the terms coming from the Wilson loops also contribute to these
integrations unlike in the case of $Z_0$ in Eqn. (\ref{z0}). 
The partition function becomes
\be
Z_{1}^\prime=\int D{\t \W}~\left[
{\Pi}_{i=1}^{\infty}2^n\int~\fr{d\b_i}{\b_i}\int Dq_{\m}^i(\tau)
~e^{-\int_{0}^{\b} d\tau\left(\fr{1}{4}(\d_{\m\n}-C_{\m\n}){\dot
q}_{\m}^i{\dot q}_{\n}^i + V\right)}\right]~\d(\chi)~e^{-S_1}
\label{partf}
\ee
where the measure is $D{\t\W}= DC D\a D\eta DB Da$. The integration
over the vector potential $A_\m$ gives the delta function in 
Eqn. (\ref{partf}). The explicit form of this delta function is
\be
\delta(\chi)\equiv\delta(F_{\m}(\theta)+iJ_\m-4g^2i\p_\a C_{\a\m}-2i\cdot{4g^2}k\e_{\m\n\l}\p_\n
a_\l)
\label{act3}
\ee
with $F_{\m}(\theta)$ as given in Eqn.(\ref{ft}). The action $S_1$ in
Eqn. (\ref{partf}) is given as
\be
S_1=S_{eff}^{\prime}-\int d^3 x~4g^2\left[
(k\e_{\m\n\l}\p_\n a_\l)^2-\fr{2k}{4g^2}\e_{\m\n\l}J_\m\p_\n a_\l+
ik\e_{\m\n\l}a_\m\p_\n a_\l+\fr{i}{8g^2}\a_\m J_\m\right].
\label{act2}
\ee
The $S_{eff}^\prime$ here is same as $S_{eff}|_{v_{\m}=0}$. The $J_\m$
that appears in Eqns. (\ref{act3})~and~(\ref{act2}) is the current
associated with the particles moving along the Wilson loops $C_i$ and
is given by 
\be 
J_\m=\Sigma_{i}^{n}\int \fr{\p q_{\m}^i}{\p
\tau}~\delta^3(q-q_{\m}^{C_i}).
\label{cur}
\ee

From Eqn. (\ref{act3}), we note that even when the co-efficient of the
Hopf term $\l$ is set to zero (i.e.,$k=0$) the current $J_\m$ do {\it not}
vanish because of the $\theta$ dependant terms. Thus the
non-commutativity of the space-time which gave rise to new interaction terms 
also results a {\it non-vanishing} current even when $\l=0$. This has to be
contrasted with the commutative case where the current vanishes when
$\l=0$ signaling the confinement of particles and anti-particles and a
non-vanishing $\l$ leads to deconfinement\cite{trg}. Here, in our case,
we see that when the NC parameter is non-vanishing, there is no
confinement of particles and anti-particles even when $\l=0$.

Now we integrate over the field $a_\m$ in the partition function
in Eqn.(\ref{partf}). With the delta function in Eqn.(\ref{act3}) this
is done trivially, leading to
\be
Z_1=\int D{\bar\W}
~\left[
{\Pi}_{i=1}^{\infty}2^n\int~\fr{d\b_i}{\b_i}\int Dq_{\m}^i(\tau)
~e^{-\int_{0}^{\b} d\tau\left(\fr{1}{4}(\d_{\m\n}-C_{\m\n}){\dot
q}_{\m}^i{\dot q}_{\n}^i + V\right)}\right]~e^{-(S_{eff}+S_j)}
\label{partz1}
\ee
where
\bea
S_j&=&-\int d^3x\left[\fr{i}{4\cdot 4g^2k}J_{\m}d_{\m\n}^{-1}J_{\n}+
\fr{1}{2\cdot 4g^2}
J_\m\left(i(F_{\m}(\theta)+4g^2\a_\m)+\fr{1}{k}d_{\m\n}^{-1}F_{\n}(\theta)
\right)\right]\no\\
&+&J_\m\left[\fr{i}{2k}d_{\m\n}^{-1}+\d_{\m\n}\right]\p_\a C_{\a\n}-
\int d^3 x\left[\fr{1}{3}v_{\m}^2+\fr{1}{2g}J_\m v_\m \right]
\eea
and $D{\bar\W}= DC D\a D\eta DB Dv$. Here the field $v_\m$ is
introduced to linearise the quadratic term in the current $J_\m$.
With $J_\m$ as given in Eqn. (\ref{cur}) and
$d_{\m\n}=\e_{\m\n\l}\p_\l$, the contribution from the first term of
$S_j$ is well known;
\be
e^{-\fr{i\pi^2}{\l}\int d^3x J_{\m}d_{\m\n}^{-1}J_{\n}}=
e^{\fr{i\pi^2}{\l}\left(\Sigma_{i=1}^n {\cW}(C_i)+\Sigma_{i\ne
j}2n_{ij}\right)}
\label{writhe}
\ee
In the above ${\cW}(C_i)$ is the writhe of the curve which in terms of the
solid angle subtended  by the tangent to the $C_i$ on a sphere traced
out by it and an {\it odd} integer as ${\cW}(C_i)=\fr{1}{2\pi}\W(C_i)+(2k+1)$.
$n_{ij}$ is the linking number of the curves $C_i$ and $C_j$ and its
contribution to partition function is {\it unity} when $\l=\fr{\pi}{2s}$.
Using these results in Eqn.(\ref{partz1}), we get
\be
Z_1=\int D{\bar\W}
~\left[
{\Pi}_{i=1}^{\infty}2^n\int~\fr{d\b_i}{\b_i}\int_{ q_{\m}}
~e^{-\int_{0}^{\b} d\tau\left(\fr{1}{4}(\d_{\m\n}-C_{\m\n}){\dot
q}_{\m}^i{\dot q}_{\n}^i + V\right)+{(-)^{2s}}is\W -iV_\m J_\m}\right]
~e^{-S_{eff}}
\label{fz1}
\ee
Here
\bea
V_\m&=&\fr{i}{2\cdot 4g^2}\left(i(F_{\m}(\theta)+4g^2\a_\m)
+\fr{1}{k}d_{\m\n}^{-1}F_{\n}(\theta)\right)
+\fr{i}{2g}v_\m+(\fr{1}{2k}d_{\m\n}^{-1}-i\d_{\m\n})\p_\a C_{\a\n}
\label{vecpot}
\eea
Notice that the $(-)^{2s}$ factor in Eqn. (\ref{fz1}) above. This
factor is due to the odd integer $2k+1$ appearing in the expression of writhe
${\cW}(C_i)$.

We now use the $Z_0$ and $Z_1$ given above in Eqn.(\ref{pf2}) to get
\be
Z=\int D{\bar\W}~e^{-\left(S_{eff} -2
\int~\fr{d\b}{\b}\int_{ q_{\m}}
~e^{-\int_{0}^{\b} d\tau\left(\fr{1}{4}(\d_{\m\n}-C_{\m\n}){\dot
q}_{\m}^i{\dot q}_{\n}^i + V\right)+{(-)^{2s}}is\W -iV_\m J_\m}\right)}
\label{dpf1}
\ee
Here we notice that the $\theta$ dependence of the partition function
comes from $S_{eff}$ and also through the potential $V_\m$.

The effect of adding the Polyakov phase factor to the path integral of
spinless particle for free as well as in presence of background
fields has been studied and it is well known to give path integral
corresponding to  particles with spin $s$ \cite{bose,trg}. This has
been shown using the $SU(2)$ coherent state path integral which gives
\be
\int_{{\hat U}(0)={\hat U}(\l)}{\cD}{\hat U} ~e^{is \int_{0}^{\l}
d{\tau}(H({\hat U})+ \W)}=Tr\left<{\hat U}| ~e^{{is}
H(\tau_\m)}|{\hat U}\right>
\ee
where ${\hat U}$ are the $SU(2)$ coherent states and $\tau_\m$ are the
generators of spin-$s$ representation of $SU(2)$\cite{peri}. We adapt
these results to our present case and obtain
\be
\int_{\L^{-2}}^{\infty}~\fr{d\b}{\b}\int_{ q_{\m}}
~e^{-\int_{0}^{\b} d\tau\left(\fr{1}{4}(\d_{\m\n}-C_{\m\n}){\dot
q}_{\m}^i{\dot q}_{\n}^i + V\right)+{(-)^{2s}}is\W -i\oint V_\m dx_\m}=
(-)^{2s}\int_{\L^{-2}}^{\infty}~\fr{d\b}{\b}
Tr~e^{-\b\left[\fr{D}{s\cA}+\fr{{\sqrt{\pi}}\e}{4}V+M\right]},
\label{det}
\ee
where $\L$ is the cut-off and $D={\rm sgn}(\l)(i\p_\m-V_\m)\tau^\m$. Here $\tau_\m$ are the
generators of spin-s representation of $SU(2)$,
$M=\L\fr{\sqrt{\pi}~\ln(2s+1)}{4}$,
${\cA}=\sqrt{\det(\d_{\m\n}-C_{\m\n})}$ and $V$ is defined in Eqn. (\ref{vp}).

Using this in Eqn.(\ref{dpf1}) we get
\be
Z=\int D{\bar\W}~e^{-S_{eff}~(-1)^{2s+1}}\det
\left[\fr{D}{s\cA}+\fr{{\sqrt{\pi}}\L^{-1}}{4}V+M\right]
\ee
The above determinant can be expressed as a functional integral over
${\bar\Psi}$ and $\Psi$ which are complex doublet fields or fermionic
fields depending whether $2s+1$ is odd or even integer. Here we see
that the factor $(-)^{2s}$ appeared in Eqn. (\ref{fz1}) (coming from the
writhe of the Wilson loop calculated in Eqn. (\ref{writhe})) is the
important factor deciding the statistics of the dual theory. This
factor of $(-)^{2s}$ in the exponential in Eqn ({\ref{fz1}), in turn, is
obtained by choosing the coefficient of the Hopf term in
Eqn. (\ref{cp1act}). Since the Hopf term do not change under the SW
map, we see that the NC parameter do not affect the statistics of the
dual fields.

Thus exponentiating the determinant in the above, we get the partition
function as
\be
Z=\int D{\bar\W}D{\bar\Psi}D{\Psi}~e^{-S_{eff}} e^{-
\int d^3 x {\bar\Psi}
\left[ \fr{D}{s\cA} +\fr{{\sqrt{\pi}}}{4}V+M\right]\Psi}
\label{dz}
\ee

Thus we see that all the dependence on the NC parameter $\theta$ comes
through terms linear and quadratic $F_{\m}(\theta)$ appearing in
$S_{eff}$ and also from the ${\bar\Psi}V_\m\tau_\m\Psi$ where it is
coupled linearly. Since we have kept only terms of order $\theta$ in SW
map while writing the NC action in terms of commutative fields and
$\theta$ in Eqn.(\ref{swcpact}), in the action ${-S_{eff}^\prime}$ appearing
in Eqn. (\ref{dz}) also we keep only linear terms in $\theta$ and
carry out integrations over $\a_\m,v_\m$ and $\eta$. Thus we get the
partition function of the dual theory as
\be
Z=\int DC DB D{\bar\Psi}D{\Psi}~e^{-{\cS}}
\label{pfd}
\ee
where the dual action is 
\bea
{\cS}&=& \int d^3 x~C_{\m\n}B_{\m\n} -\fr{1}{k}\p_\n C_{\m\n}
d_{\m\a}^{-1}\left[ig^2\p_\b
C_{\b\a}-\fr{1}{2}F_{\a}(\theta)\right]
-\fr{i}{2}F_{\m}(\theta)\p_\a C_{\a\m} +g^2(\p_\n C_{\m\n})^2\no\\
&+&g^2 \p_\a C_{\m\a}(\d_{\m\n}+C_{\m\n}^{-1})\p_\b C_{\b\n}
+{\bar\Psi} \left[\fr{{\rm sgn}(\l)}{s{\cA}}
(i\p_\m-{\widetilde V}_\m)\tau^\m+(M+2g^2\r\sqrt{\pi}{\L}^{-1}~)\right]\Psi
\no\\
&+&\fr{\pi\r}{16\L}({\bar\Psi}\Psi)^2
+\fr{3}{16g^2{s^2\cA}^2}({\bar\Psi}\tau_\m\Psi)^2
+\fr{1}{16 g^2s^2{\cA}^2}({\bar\Psi}\tau_\m\Psi)(\d_{\m\n}+C_{\m\n}^{-1})^{-1}
({\bar\Psi}\tau_\n\Psi)
\label{dact}
\eea
Here
\be
{\widetilde V}_\m= \fr{i}{8g^2k}d_{\m\n}^{-1}F_\n(\theta)+
\fr{1}{2}(\fr{1}{k}d_{\m\n}^{-1}-i\d_{\m\n}+ C_{\m\n}^{-1})\p_\a C_{\a\n} 
\ee
Notice here that through ${\widetilde V}_\m$, the $\theta$ dependant 
terms directly get coupled to $\bar\Psi$ and $\Psi$. The dual action
has further $\theta$ dependant terms which are coupled to the
auxiliary field $C_{\m\n}$. In the commutative limit all these later
terms vanishes and the integrations over the fields $B_{\m\n}$ and
$C_{\m\n}$ becomes trivial giving the action in the commutative 
plane 
\be
S=\int d^3x~ {\bar\Psi} \left[\fr{{\rm sgn}(\l)}{s}
i\tau^\m \p_\m+(M+2g^2\r\sqrt{\pi}{\L}^{-1}~)\right]\Psi
+\fr{1}{4g^2s^2}({\bar\Psi}\tau_\m\Psi)^2
+\fr{\pi\r}{16\L^2}({\bar\Psi}\Psi)^2
\ee
obtained in \cite{trg}. The four-fermi interaction term in the dual theory
comes when we integrate over the auxiliary field $\eta$.  This field
was introduced in the action (see Eqn. (\ref{sact})) to incorporate the
condition in Eqn.(\ref{mod}) which is the same in the commutative case
also. Thus it is not surprising to see that the four-fermi interaction
term in both NC case and commutative model are the same. In contrast,
the Thirring term gets a $\theta$ dependence (through ${\cA}$). Notice
that the duality shown here is exact, to all orders in fermion mass
and coupling constants.

Our dual theory in terms of spin-s fields is non-local as expected for
a field theory on a noncommutative space. Our first aim is to see what
is the dual theory for a $CP^1$ model in NC spacetime by starting from
a SW mapped $CP^1$ model. Our results clearly point to the fact that
the dual action obtained here (Eqn. \ref{dact}) is not the na\"ive NC
generalisation of the commutative action obtained in \cite{trg} (but
in the limit $\theta\to 0$, we recover the action obtained in
\cite{trg}). Similar feature was also noticed in the context of the
duality between Maxwell-Chern-Simons theory and self-dual model in the
NC settings\cite{sg}. Also it has been shown that the effect of
NC is same as that of a field dependent gravitational
background\cite{vor} and thus the proper time determinant in
Eqn. (\ref{popdet}) can be thought of as evaluated in a non-trivial back
ground. It is this background dependence coming bacuse of the
non-commutativity which leads to the appearence of ${\cal A}^{-1}$ and
$(\d_{\m\n}+C_{\m\n}^{-1})^{-1}$ in the dual action. Inspite of these
nonlocal and non-polynomial nature of the dual theory one would be
able to show various relations between these theories by formally
taking functional derivatives.

From the equivalence of the partition functions in Eqn. (\ref{pfe}) and
Eqn. (\ref{pfd}) obtained here we can derive the mappings between
various n-point correlators of $CP^1$ model and the dual spin-$s$
theory in the NC plane by introducing appropriate source terms. The
form of the SW mapped Hopf term and the SW mapped field strength of
the vector field is suggestive to couple a topologically invariant
current of the form
\be
J_{\m}^{top}=\fr{1}{2\pi}\e_{\m\n\l}\p_\n A_\l
\ee
using a source (a vector field here) to the partition function of the
SW mapped $CP^1$ model with Hopf term in Eqn.(\ref{pfe}). Repeating the
steps leading to Eqn.(\ref{pfd}), we get the dual partition function
where the source filed dependant term are present. Now by taking
functional derivatives we get
\be
\left<J_{\m}^{top}\right>_{NC CP^1}=2s i\left<J_{\m}^{N}+ 2
F_{\m}(\theta)-i4g^2\p_\n C_{\m\n}+\fr{1}{8s\pi}\e_{\m\n\l}\p_\l\p_\a
C_{\a\n}\right>_{NC spin-s}
\ee
where $J_{\m}^{N}=sgn{\l}~{\bar\Psi}\fr{\tau_\m}{s{\cA}}\Psi$. The
over all factor $i$ in the above will be removed when we do a Wick
rotation from Euclidean space. From Eqn. (\ref{dact}) it is clear that
the current $J_{\m}^{N}$ gets the $\theta$ dependence through ${\cA}$.
Thus the above map between correlators shows the interesting feature that the
Noether charge is $\theta$ dependant where as the corresponding
soliton charge is not. We also notice that a spin-$s$ particle in the
NC dual theory corresponds to a soliton of index $2s$ as in the
commutative case.

\section{Conclusion}

We have studied the duality equivalence in the NC plane and showed
that the NC $CP^1$ model coupled with Hopf term is equivalent to an
interacting NC spin-$s$ theory with $s=\fr{\pi}{2\l}$ where $\l$ is
the coupling strength of the Hopf term. We have shown this
equivalence after re-expressing the NC $CP^1$ model with Hopf term
using SW map, keeping terms to order $\theta$. We recover the dual
interacting spin-$s$ theory constructed in the commutative plane in
the limit $\theta\to 0$ from the NC dual theory obtained here. 
There are couple of points worth mentioning here. Ours is among the 
first to study the NC dual equivalence using path integral approach.
Secondly dual of $CP^1$ model in NC space is different from NC version of
the dual of $CP^1$ model in commutative space.
We have also shown here that the statistics of the dual theory do not get
affected by the non-commutativity of the space time. The mapping
between the correlators of topological and Noether currents shows that
while the topological index is unaffected by NC parameter $\theta$ the
Noether charge of the NC dual theory depends on $\theta$.

It will be of interest to see what are the new solutions in NC $CP^1$
model obtained in \cite{cp1} corresponds to in the dual spin-$s$
theory obtained here.  $CP^1$ model coupled with Hopf term has been
constructed and studied in the non-commutative sphere
also\cite{trghk}. It will be interesting to study whether the
equivalence obtained here can be generalised to fuzzy sphere and to
analyse the various limits of the dual theory on fuzzy sphere.

\nn{\bf ACKNOWLEDGEMENTS}:
We thank Subir Ghosh for useful discussions regarding the validity of
the use of commutative $CP^1$ constraint in the SW mapped theory. EH
thank V. O. Rivelles for useful discussions. He also thank FAPESP for
support through grant 03/09044-9.

\end{document}